\documentclass{article}
\usepackage{amssymb}

\usepackage{graphicx}
\usepackage{amsmath}
\usepackage{Latexcad}

\input{tcilatex}

\begin{document}

\title{Anomalous Scale Dimensions from Timelike Braiding}
\author{Bert Schroer \\
presently: CBPF, Rua Dr. Xavier Sigaud, 22290-180 Rio de Janeiro, Brazil \\
email: schroer@cbpf.br\\
Prof. emeritus of the Institut f\"{u}r Theoretische Physik\\
FU-Berlin, Arnimallee 14, 14195 Berlin, Germany}
\date{September, 2000}
\maketitle

\begin{abstract}
Using the previously gained insight about the particle/field relation in
conformal quantum field theories which required interactions to be related
to the existence of particle-like states associated with fields of anomalous
scaling dimensions, we set out to construct a classification theory for the
spectra of anomalous dimensions. Starting from the old observations on
conformal superselection sectors related to the anomalous dimensions via the
phases which appear in the spectral decomposition of the center of the
conformal covering group $Z(\widetilde{SO(d,2)}),$ we explore the
possibility of a timelike braiding structure consistent with the timelike
ordering which refines and explains the central decomposition. We regard
this as a preparatory step in a new construction attempt of interacting
conformal quantum field theories in D=4 spacetime dimensions. Other ideas of
constructions based on the $AdS_{5}$-$CQFT_{4}$ or the perturbative SYM
approach in their relation to the present idea are briefly mentioned.
\end{abstract}

\section{Background and Preview of new Results}

It had been known for a long time that conformal quantum field theory
exhibits in addition to the general spin-statistics theorem another more
characteristic structural property which we will refer to as the ``anomalous
dimension-central phase'' connection. It relates the anomalous scale
dimension of fields modulo integers (semi-integers in the case of Fermion
fields) to the phase obtained by performing one complete timelike sweep
around the compactified Minkowski world \cite{S-S} and hence is analogous to
the univalence superselection rule of the semi-integer spin which
historically spells the beginning of the issue of superselection rules in
the famous paper of Wick, Wightman and Wigner \cite{WWW}. In the spin case
one associates with a $2\pi $ spatial rotation sweep the statistics phase $%
\left( -1\right) ^{2s}$ of the spin-statistics connection \cite{St-Wi} and
the question of whether there is a commutation relation behind the
superselected coherent subspace corresponding to each phase factor of the
timelike sweep poses itself naturally. The word ``central'' here refers to
the center $Z$($\widetilde{SO(d,2)}$) of the infinite sheeted covering group 
$\widetilde{SO(d,2)}$ which has one abelian generator $Z$ for spacetime
dimensions $d>1+1.$ In chiral conformal theories the sweep is lightlike and
the spin and its timelike analogue coalesce, whereas in higher dimensions
there is the problem of consistency with the well known DHR superselection
theories of internal symmetries based on spacelike commutativity.

It is our aim to show that the analogy is deeper than expected at first
sight, namely the statistics aspect of the spin has an algebraic counterpart
in form of a timelike braid group (''plektonic'') commutation relation. Here
the notion of global causality in the covering of the compactified Minkowski
spacetime is important because it was on the basis of this concept that the
``Einstein causality paradox'' \cite{HSS} was solved \cite{S-S}. The
conformal decomposition theory resulted from the attempt to avoid the
covering formalism (which is not natural from a particle physics viewpoint)
and to deal instead with projected fields which behave like sections over
compactified Minkowski spacetime rather than globally causal fields on the
covering. Whereas the latter are Wightman fields, the former are not since
they carry with them a source and a range projector\footnote{%
Such operators are sometimes called ''vertex operators''. The old name
``central component- or projected fields'' \cite{S-S} or the more recent
name ``exchange algebra fields'' suits better the content of the present
paper.}, and hence are similar to the exchange algebra fields of chiral
theories \cite{R-S}. As in the chiral case the spectrum of anomalous
dimensions (possibly modulo a common abelian contribution) is determined in
terms of the admissable braid group representations. The chiral observables
on $S^{1}$ correspond to conformal observables on compactified Minkowski
space $\bar{M}$. For the latter one has space- and timelike commutativity
(validity of Huygens principle), in fact this distinction and the notion of
causality altogether becomes meaningless and only lightlike distances have
an invariant meaning.

In d=1+1 dimensions one has accumulated a good understanding of conformal
theories and in particular of their associated superselected charge
structure. One knows that they can be decomposed into the $x_{\pm }$ chiral
light cone components. There is a systematic way to classify localizable
representation of chiral observable algebras (at least in principle) and one
finds charge-carrying fields which obey a lightlike exchange algebra \cite
{RS Ising}\cite{R-S} in which those new objects satisfy braid group
commutation relations either of the abelian kind (anyonic) or with plektonic
(which includes the nonabelian case) R-matrices with quantized statistical
phases. Since the latter determine the spectrum of anomalous dimensions (or
spectrum of ``twists''= scale dimension minus spin) modulo integers$,$ one
has a theory of anomalous dimension as soon as one knows how to classify
physically admissable representations of the infinite braid group or more
precisely the ribbon braid group $RB_{\infty }$. The classification of the
latter is done by the method of tracial states on $B_{\infty }$ which follow
a combinatorial version of the field theoretic cluster decomposition
property, the so-called Markov property \cite{Jones}\cite{Wenzl}. This
method was originally invented in the early 70$^{ies}\,$by Doplicher Haag
and Roberts (DHR) in order to classify the admissable permutation group
statistics which is associated with the algebraic superselection theory of
compactly localized charges in $d>1+1$ \cite{Haag} in a formulation without
field multiplets. In subfactor theory \cite{GHJ} this method was
independently discovered in a vastly more general context and called it very
appropriately the method of ``Markov traces'' (due to Vaughn Jones) which in
turn was gratefully re-adopted by the physicists. The name Markov in this
context reveals a lot about the conceptual scope of this theory because
Markov junior refers to the Russian mathematician who made important
contribution to the early study of the braid group, but at the same time one
is invited to think about Markov senior the probabilist since, while for a
physicist the property of this tracial state (which then allowed its
iterative determination) was a discrete version of the field theoretic
cluster decomposition property, to a mathematician this procedure was more
reminiscent of a discrete stochastic process. A field theoretic version of
the classification of admissable braid group representations based on the
Markov trace formalism can be found in \cite{FRSI}\cite{FRSII}\cite{Paler} 
\cite{subfac}

The reader will notice that in our enumeration of achievements in chiral
theories we have omitted the better known representation theory of specific
algebras as the energy-momentum tensor algebra or current algebras. This is
not to ignore their important role in the modern development of chiral
theories but rather a result of the fact that they have no direct
counterpart in higher dimensions. So if we want to use chiral theory as a
theoretical laboratory for higher dimensional conformal field theories we
are forced to de-emphasized those aspects and highlight instead others, as
space- and time-like commutation structures which are independent on
spacetime dimensions.

The weak point of the present approach to higher dimensional conformal
theories is of course the total lack of nontrivial examples i.e. of higher
dimensional conformal models with anomalous scaling dimensions. As already
mentioned the new timelike superselection structure can unfortunately not be
illustrated by the representation theory of any known algebra (unlike chiral
theory) and it is also not possible to explore this timelike region by
Lagrangian methods (which tend to favor the euclidean or spacelike regions).
There is of course the folklore that the only Lagrangian conformal
4-dimenional theories are a special kind of supersymmetric Yang-Mills
theories which, if it could be made more rigorous in a clearer conceptional
setting (less computational recipes, more physical principles) would be a
remarkable observation. The present non-Lagrangian approach suggests another
picture: instead of a scarceness of models one should expect a similar
wealth as in the case of nonperturbative chiral theories. I think that in
the near future one will have new concepts and methods for their
construction from timelike braid group data. More remarks can be found at
the end.

In lack of illustrative examples which could show that the present
requirements allow for nontrivial realizations, we are limited to
consistency checks. The main new problem which was absent in the chiral case
is the consistency of timelike braiding with spacelike locality i.e. one has
to show that the new timelike plektonic structure is in harmony with the
standard Boson/Fermion local commutation relations for the spacelike region.
This will be the subject of the third section.

The next section contains a review of the geometric setting of conformal
symmetry. Although most of these results are known since the 70s, we find it
convenient for setting notation and concepts to present an updated version.
The isomorphism of conformal field theory with QFT in 5-dim. anti de Sitter
spacetime belongs logically (though not historically) to that section which
deals with issues of compactification, covering and global causality.

Some comments on the relation of conformal QFT and particle physics are in
order. In most of the recent literature standard notions of particle physics
as particles and scattering theory have been used in the conformal setting
without qualifications . But a closer examination shows that conformal field
theory is not a theory of interacting particles, at least not if these
concepts are still used in a way which is not completely void of their
original physical meaning. There are no discrete zero mass shells (light
cone Delta functions in momentum space) in an \textit{interacting} dilation
invariant theory and there is a fortiori no LSZ S-matrix (the asymptotic LSZ
limits in fact vanish). Any kind of interaction dissolves immediately the
zero mass shell into the continuum with a continuous mass distribution of
enhanced weight at $p^{2}=0$. There remains of course the interesting
question of what kind of residual particle physics information\footnote{%
Although scattering amplitudes diverge or go to zero in the scaling limit,
highly inclusive cross sections could stay finite.} one can extract from the
scale invariant limit of a massive particle theory. For a discussion of this
and related points see \cite{particle}. The structural problems of conformal
QFT to which we draw our attention in the sequel do not depend on particle
interpretations.

\section{Covering Space and Decomposition Theory}

Since massless particles in a conformal theory cannot interact \cite
{particle}, the physically interesting interacting fields in a conformal
theory are those with anomalous dimension. Whereas conformal free fields are
commutative in the Huygens sense of time- and spacelike commutativity
(Huygens principle in wave optics transposed into the setting of local
quantum physics), conformal interactions lead to ``reverberations$"$ inside
the light cone which are in correspondence with the appearance of anomalous
dimension and produce a paradoxical violation \cite{HSS} of Einstein
causality if one uses the standard description of Minkowski space and allows
``big'' special conformal transformations to act on spacelike vanishing
commutators with the usual transformation rules. Hence it was of interest to
investigate the global spacetime interpretation of such fields in more
detail in order to remove the paradox. For aspects of global localization in
interacting conformal field theories one needs to introduce the covering of
the conformally compactified Minkowski space with its nontrivial topology.
This is a well-studied old subject \cite{Se}\cite{S-S}\cite{LM}, which led
to an important global decomposition theory which, after lying dormant for a
number of years (when gauge theories took the center of the stage), \cite
{S-V}obtained an unexpected fresh push in the special case of d=1+1 \cite
{BPZ} where conforms observables decompose (similar to free fields)
additively in to the two light cone components. The latter act on a
tensor-factorized Hilbert space whereas charge-carrying fields act on a
tensor product of extended superselected Hilbert spaces.

The fastest way to obtain a first glance at the formalism and physical use
of the conformal covering space is to notice that the Wigner representation
theory for the Poincar\'{e} group for zero mass particles allows an
extension to the conformal symmetry (without extending the Hilbert space and
the degree of freedoms): Poincar\'{e} group $\mathcal{P}(d)\rightarrow
SO(d,2)$. Besides scale transformations, this larger symmetry also
incorporates the fractional transformations (proper conformal
transformations) 
\begin{eqnarray}
x^{\prime } &=&\frac{x-bx^{2}}{1-2bx+b^{2}x^{2}}=IT(b)I  \label{sc} \\
I &:&x\rightarrow \frac{-x}{x^{2}},\,\,\,T(b):x\rightarrow x+b  \notag
\end{eqnarray}
The conformal reflection $I$ itself is not a Moebius transformation, but in
free field theories it is known to be implemented by a unitary
transformation \cite{S-V}. For fixed x and small b the formula (\ref{sc}) is
well defined, but globally it mixes finite spacetime points with infinity
and hence requires a more precise definition (in particular in view of the
positivity energy-momentum spectral properties) in its action on quantum
fields.

As preparatory step for the quantum field theory concepts one has to achieve
a geometric compactification. This starts most conveniently from a linear
representation of the conformal group SO(d,2) in 6-dimensional auxiliary
space $\mathbb{R}^{(d,2)}$ (i.e. without field theoretic significance) with
two negative (time-like) signatures 
\begin{equation}
G=\left( 
\begin{array}{ccc}
g_{\mu \nu } &  &  \\ 
& -1 &  \\ 
&  & +1
\end{array}
\right) ,\,\,g=diag(1,-1,-1,-1)
\end{equation}
and restricts this representation to the (D+1)-dimensional forward light
cone 
\begin{equation}
LC^{(d,2)}=\left\{ \xi =(\mathbf{\xi },\xi _{4},\xi _{5}\right\} ;\mathbf{%
\xi }^{2}+\xi _{d}^{2}-\xi _{d+1}^{2}=0\}
\end{equation}
where $\mathbf{\xi }^{2}=\xi _{0}^{2}-\vec{\xi}^{2}$ denotes the
D-dimensional Minkowski length square. The compactified Minkowski space is
obtained by adopting a projective point of view (stereographic projection) 
\begin{equation}
M_{c}^{(d-1,1)}=\left\{ x=\frac{\mathbf{\xi }}{\xi _{d}+\xi _{d+1}};\xi \in
LC^{(d,2)}\right\}
\end{equation}
It is then easy to verify that the linear transformation which keep the last
two components invariant consist of the Lorentz group and those
transformations which only transform the last two coordinates yield the
scaling formula 
\begin{equation}
\xi _{d}\pm \xi _{d+1}\rightarrow e^{\pm s}(\xi _{d}\pm \xi _{d+1})
\end{equation}
leading to $x\rightarrow \lambda x,\lambda =e^{s}\,.$ The remaining
transformations, namely the translations and the fractional proper conformal
transformations, are obtained by composing rotations in the $\mathbf{\xi }%
_{i}$-$\xi _{d}$ and boosts in the $\mathbf{\xi }_{i}$-$\xi _{d+1}$ planes.

The so obtained spacetime is most suitably parametrized in terms of a
``conformal time'' $\tau $%
\begin{align}
M_{c}^{(d-1,1)}& =(sin\tau ,\mathbf{e,}cos\tau ),\,\,e\in S^{3}  \notag \\
t& =\frac{sin\tau }{e^{d}+cos\tau },\,\,\vec{x}=\frac{\vec{e}}{e^{d}+cos\tau 
}  \label{M} \\
e^{d}+cos\tau & >0,\,\,-\pi <\tau <+\pi   \notag
\end{align}
so that the conformally compactified Minkowski space is a piece of a
multi-dimensional cylinder carved out between two d-1 dimensional boundaries
which lie symmetrically around $\tau =0,\mathbf{e=(0},e^{d}=-1\mathbf{)}$
where they touch each other \cite{LM} (Fig.1 below). If one cuts the
cylinder wall this region $\bar{M}^{(d-1,d)}$ looks like a double cone
subtended by two points at infinity $m_{+}(\tau =\pi ,\vec{e}=0,e^{d}=1),$ $%
m_{-}(\tau =-\pi ,\vec{e}=0,e^{d}=1);$ the boundary region $\bar{M}%
\backslash M$ consists precisely of all points which are past/future light
like from $m_{+}/m_{-}$ (the light rays on the cylinder are continuous lines
starting from $m_{-}$ and ending at $m_{+}$ which are points to be
identified on $\bar{M}$). In this way the cylinder is equipped with a tiling
into infinitely many ordinary Minkowski spaces. 
\begin{equation}
\widetilde{M_{c}^{(d-1,1)}}=S^{d-1}\times \mathbb{R}
\end{equation}
Whereas matter in $\bar{M}$ is subject to the quantum version of the Huygens
principle namely observables commute if the light ray subtended by one
localization region cannot reach the other, the covering $\widetilde{M}$
comes with a global causality structure. The relevance of this covering
space for the notion of relativistic causality was first pointed out first
by I. Segal \cite{Se} and the above parametrization as well as many other
contributions which became standard in conformal QFT appeared for the first
time in the work of Luescher and Mack \cite{LM}.

Formally this framework solves the ``Einstein causality paradox of conformal
quantum field theory'' \cite{HSS} which originated from ``would be''
conformal models (locally conformal invariant) of quantum field theory as
e.g. the massless Thirring model which violates Huygens principle. The naive
reason for this apparent violation turned out to be that there exist
continuous curves of special conformal transformations which lead from
timelike separations with one point at the origin via the lightlike infinity
to spacelike separation. This obviously generates a contradiction with the
locality structure of the Thirring model whose timelike anti-commutator
unlike the spacelike one does not vanish. The covering structure formally
solves this causality paradox by emphasizing that in fact there are no
covering transformation which violate global causality, a paradox only
arises if points become projected out of $\widetilde{M}$ into $M.$ In other
words, one must keep track of the global path which remains space/time- like
and not just its end points. If one depicts as before the covering space as
a cylinder (Fig.1 below), then it contains infinitely many copies of the
original Minkowski space which appear in the projection to ($\xi _{2},..\xi
_{d-1})=(0,..0)$ subspace as a finite rhomboid region \cite{LM}.

Using the above parametrization in terms of $\mathbf{e}$ and the ``conformal
time'' $\tau ,$ one can immediately globalize the notion of time like
distance and one finds the following causality structure (\cite{Se}\cite{LM}%
) 
\begin{align}
& \left( \xi (\mathbf{e},\tau )-\xi (\mathbf{e}^{\prime },\tau ^{\prime
})\right) ^{2}\lessgtr 0,\,\,hence  \label{glob} \\
& \left| \tau -\tau ^{\prime }\right| \lessgtr \left| 2Arcsin\left( \frac{%
\mathbf{e}-\mathbf{e}^{\prime }}{4}\right) ^{\frac{1}{2}}\right| =\left|
Arccos\left( \mathbf{e\cdot e}^{\prime }\right) \right|  \notag
\end{align}
where $\lessgtr $ now denotes the spacelike/$\left( \pm \right) $timelike
separation in the global sense. Since it is expressed in terms of the
difference of two coordinates on the light cone in the 6-dim. ambient
spacetime, a conformal transformation which is linear in the $\xi $%
-variables leaves it invariant. For the description of the Dirac-Weyl
compactified Minkowski space the use of the following simpler
parametrization close to standard Minkowski coordinates is more convenient 
\begin{eqnarray}
&&\xi ^{\mu }=x^{\mu } \\
&&\xi ^{4}=\frac{1}{2}(1+x^{2})  \notag \\
&&\xi ^{5}=\frac{1}{2}(1-x^{2})  \notag \\
&&i.e.\,\,\,\left( \xi -\xi ^{\prime }\right) ^{2}=-\left( x-x^{\prime
}\right) ^{2}  \notag
\end{eqnarray}
Similarly one may use the quadratic polynomial $\sigma (b,x)$ appearing in
the denominator (and the Jacobian) of the special conformal transformations
in order to decide whether two points are globally timelike/spacelike
(connectable by timelike/spacelike geodesics) without using (\ref{glob}) 
\cite{S-S}\cite{Haag}.

The formulation in terms of conformal covering space would be useful if the
world (including laboratories of experimentalists) would also be conformal,
which certainly is not the case. Therefore it is helpful to know that there
is a way of re-phrasing the physical content of local fields (which violate
the Huygens principle and instead exhibit the phenomenon of
``reverberation'' \cite{HSS} inside the forward light cone) in the Minkowski
world $M$ of ordinary particle physics without running into the trap of the
causality paradox of the previous section; in this way the use of the above $%
\xi $- parametrization would loose some of its importance and this may be
considered as an alternative to the L\"{u}scher-Mack approach on covering
space.

This was first achieved in a joint paper involving one of the present
authors \cite{S-S} whose main point was that the global causality structure
could be encoded into a global decomposition theory of fields with respect
to the center of the conformal covering (conformal block decomposition).
Local fields, although behaving apparently irreducibly under infinitesimal
conformal transformations, transform in general reducibly under the action
of the global center of the covering $Z(\widetilde{SO(d,2)}).$ This
reduction was precisely the motivation in for the global decomposition
theory of conformal fields in \cite{S-S}. 
\begin{eqnarray}
F(x) &=&\sum_{\alpha ,\beta }F_{\alpha ,\beta }(x),\,\,F_{\alpha ,\beta
}(x)\equiv P_{\alpha }F(x)P_{\beta }  \label{comp} \\
Z &=&\sum_{\alpha }e^{2\pi i\theta _{\alpha }}P_{\alpha }  \notag
\end{eqnarray}
These component fields behave analogous to trivializing sections in a fibre
bundle; the only memory of their origin from an operator on covering space
is their quasiperiodicity 
\begin{eqnarray}
ZF_{\alpha ,\beta }(x)Z^{\ast } &=&e^{2\pi i(\theta _{\alpha }-\theta
_{\beta })}F_{\alpha ,\beta }(x) \\
U(b)F_{\delta }(x)_{\alpha ,\beta }U^{-1}(b) &=&\frac{1}{\left[ \sigma
_{+}(b,x)\right] ^{\delta -\zeta }\left[ \sigma _{-}(b,x)\right] ^{\zeta }}%
F_{\delta }(x)_{\alpha ,\beta } \\
\zeta &=&\frac{1}{2}(\delta _{F}+\theta _{\beta }-\theta _{\alpha })  \notag
\end{eqnarray}
where the second line is the transformation law of special conformal
transformation of the components of an operator $F$ with scale dimension $%
\delta _{F}$ sandwiched between superselected subspaces $H_{\alpha }$ anf $%
H_{\beta }.$ Using the explicit form of the conformal 3-point function it is
easy to see that phases are uniquely given in terms of the scaling
dimensions $\delta $ which occur in the conformal model \cite{S-S}. 
\begin{equation}
e^{2\pi i\theta }\in \left\{ 
\begin{array}{c}
\{e^{2\pi i\delta }|\,\delta \in scaling\,spectrum\}\,\,\,\,Bosons \\ 
\{e^{2\pi i(\delta +\frac{1}{2})}|\,\delta \in
scaling\,spectrum\}\,\,\,Fermions
\end{array}
\right.
\end{equation}
A central projector projects onto the subspace of all vectors which have the
same scaling phase i.e. onto a conformal block associated with the center,
so the labelling refers to (in case of Bosons) the anomalous dimensions $%
mod(1).$ These subspaces of operators are bigger than the chiral blocks,
since the anomalous dimension e.g. does not distinguish between with charged
fields and their anti-fields which carry the conjugate charge. Having
understood the physical interpretation of the central decomposition does not
yet mean that we have a theory of the spectrum of admissable anomalous
dimensions (critical indices) in higher dimensional conformal theories, but
the close analogy to chiral theories gives sufficient incentive to look for
such a theory. This will be the subject of the next section.

The prize one has to pay for this return to the realm of particle physics on 
$M$ in terms of component fields (\ref{comp}) is that these projected fields
are not Wightman fields. They depend on a source and range projector and if
applied to a vector the source projector has to match the Hilbert space i.e. 
$F_{\alpha ,\beta }$ annihilates the vacuum if $P_{\beta }$ is not the
projector onto the vacuum sector. This is very different from the behavior
of the original $F$ which, in case it was localized in a region with a
nontrivial spacelike complement, can never annihilate the vacuum. This kind
of projected fields are well known from the exchange algebra formalism of
chiral QFT \cite{R-S} but they appear in a rudimentary form already in \cite
{S-S}.

Inside two and 3-point functions the projectors are unique and may be
omitted but for $n\geq 4$ there are several projected n-point functions and
therefore they are needed.

Structural properties of the real time formulation as this timelike
decomposition formula remain totally hidden in the euclidean formulation.
They lead to cuts with multi-valuedness in an analyticity region which is
beyond the standard BHW \cite{St-Wi} extended permuted tube region (see next
section) of standard Poincar\'{e} invariant theories. As in the chiral case
one defines conformal observable fields as those which commute with the
center generator $Z$ and, as a consequence are free of these cuts i.e. have
meromorphic (rational) correlation functions on $\bar{M}$ and its complex
extensions as it is well known from the chiral conformal observables (where
the terminology ``holomorphic'' was unfortunately attributed to the
observable fields\footnote{%
There are no holomorphic local fields in QFT, not even for free fields. The
correlation functions have analytic properties which depend sensitively on
the state in which the correlations are studied e.g. they are very different
in ground states than in thermal states, although the local operator
algebras remain the same.} instead of referring to a particular state their
ground state correlation functions only)

The structure of the center in chiral conformal field theories is determined
by the discrete spectrum of the rotation operators for the compactified $\pm 
$lightrays $R^{(\pm )}=L_{0}^{(\pm )},$ where the right hand side is the
standard Virasoro algebra notation. It is well-known that this operator
shares with the light ray translations $P^{(\pm )}$ the positivity of its
spectrum. This becomes in fact obvious if one represents it in terms of $P$%
\begin{eqnarray}
R^{(\pm )} &=&P^{(\pm )}+K^{(\pm )} \\
K^{(\pm )} &=&I^{(\pm )}P^{(\pm )}I^{(\pm )}  \notag
\end{eqnarray}
where $I_{\pm }$ is the representer of the chiral conformal reflection $%
x\rightarrow -\frac{1}{x}$ (in linear lightray coordinates $x$) and $K$ is
the generator of the fractional special conformal transformation (\ref{sc}).
However the two-dimensional inversion does not factorize since the chiral
inversion rewritten in terms of 2-dim. vector notation corresponds to 
\begin{eqnarray}
x_{0} &\rightarrow &-\frac{x_{0}}{x^{2}} \\
x_{1} &\rightarrow &\frac{x_{1}}{x^{2}}  \notag
\end{eqnarray}
The ``wrong'' sign in the spatial part can be corrected by a parity
transformation $x_{+}\leftrightarrow x_{-}$ which mixes the two chiral
components. In defining an object which transforms as a vector this has to
be taken in consideration 
\begin{eqnarray}
R_{\mu } &=&P_{\mu }+IP_{\mu }I  \label{vec} \\
I &=&parity\cdot inversion
\end{eqnarray}

The vector formula (\ref{vec}) is valid in any dimension i.e. does not
require light ray factorization. It leads to a family of operators with
discrete spectrum $e\cdot R$ which are dependent on a timelike vector $%
e_{\mu }.$ As in the chiral case one only needs to add to the
Poincar\'{e}+scale transformations the (timelike) conformal rotation $R_{0%
\text{ }},$ the other components of $R_{\mu }$ are generated by the action
of the Lorentz group.

To understand the geometric action of $e^{ie\cdot R\tau },$ it is helpful to
depict the covering world $\widetilde{M}$ with a copy of the Minkowski world
inside. From (\ref{M}) one obtains the identification of the covering world
with the surface of a d+1 dimensional cylinder \cite{LM}. In Fig.1 only two
of the d-2 components of the d-dimensional \textbf{e}-vector have been
drawn, the others have been set zero. For depicting the spacelike complement
of a double cone $\mathcal{O}$ in in $\widetilde{M}$ it is more convenient
to cut open the cylinder in $\tau $-direction and identify opposite sides as
in Fig.2.

\begin{picture}(132,226)
\thinlines
\drawellipse{58.0}{200.0}{100.0}{20.0}{}
\drawellipse{58.0}{60.0}{100.0}{20.0}{}
\drawpath{8.0}{200.0}{8.0}{60.0}
\drawpath{108.0}{200.0}{108.0}{60.0}
\path(34.0,178.0)(34.0,178.0)(35.88,177.38)(37.75,176.82)(39.59,176.25)(41.38,175.69)(43.15,175.13)(44.88,174.6)(46.56,174.08)(48.22,173.55)
\path(48.22,173.55)(49.86,173.05)(51.45,172.55)(53.02,172.08)(54.54,171.58)(56.04,171.11)(57.52,170.67)(58.95,170.23)(60.36,169.79)(61.74,169.36)
\path(61.74,169.36)(63.08,168.94)(64.4,168.51)(65.68,168.11)(66.94,167.73)(68.18,167.33)(69.37,166.94)(70.55,166.58)(71.69,166.19)(72.83,165.83)
\path(72.83,165.83)(73.93,165.5)(75.0,165.14)(76.05,164.8)(77.05,164.48)(78.05,164.14)(79.01,163.8)(79.98,163.51)(80.9,163.19)(81.8,162.88)
\path(81.8,162.88)(82.68,162.58)(83.51,162.29)(84.37,161.98)(85.16,161.69)(85.94,161.41)(86.73,161.13)(87.48,160.86)(88.19,160.58)(88.91,160.3)
\path(88.91,160.3)(89.58,160.05)(90.26,159.76)(90.91,159.51)(91.54,159.26)(92.15,159.0)(92.73,158.75)(93.3,158.48)(93.87,158.23)(94.41,157.98)
\path(94.41,157.98)(94.94,157.73)(95.44,157.48)(95.94,157.23)(96.43,156.98)(96.9,156.75)(97.33,156.5)(97.79,156.25)(98.19,156.0)(98.62,155.76)
\path(98.62,155.76)(99.01,155.51)(99.4,155.26)(99.76,155.01)(100.12,154.76)(100.48,154.5)(100.8,154.25)(101.12,153.98)(101.44,153.73)(101.76,153.47)
\path(101.76,153.47)(102.05,153.19)(102.33,152.94)(102.62,152.66)(102.9,152.39)(103.16,152.11)(103.41,151.83)(103.66,151.55)(103.91,151.26)(104.12,150.98)
\path(104.12,150.98)(104.37,150.69)(104.58,150.38)(104.8,150.08)(105.01,149.76)(105.22,149.44)(105.41,149.14)(105.62,148.8)(105.8,148.48)(106.01,148.14)
\path(106.01,148.14)(106.19,147.8)(106.37,147.44)(106.55,147.08)(106.73,146.73)(106.91,146.36)(107.08,145.98)(107.26,145.61)(107.44,145.22)(107.62,144.8)
\path(107.62,144.8)(107.8,144.41)(107.98,144.0)(108.0,144.0)
\path(34.0,178.0)(34.0,178.0)(33.34,177.75)(32.7,177.52)(32.04,177.27)(31.44,177.05)(30.81,176.83)(30.21,176.61)(29.62,176.38)(29.04,176.16)
\path(29.04,176.16)(28.46,175.97)(27.92,175.75)(27.37,175.55)(26.81,175.35)(26.29,175.13)(25.77,174.94)(25.26,174.75)(24.76,174.57)(24.27,174.38)
\path(24.27,174.38)(23.78,174.19)(23.29,174.0)(22.84,173.8)(22.37,173.63)(21.94,173.44)(21.5,173.26)(21.05,173.11)(20.63,172.92)(20.22,172.76)
\path(20.22,172.76)(19.84,172.58)(19.43,172.41)(19.04,172.26)(18.68,172.08)(18.31,171.92)(17.93,171.76)(17.59,171.58)(17.25,171.44)(16.89,171.29)
\path(16.89,171.29)(16.56,171.11)(16.26,170.97)(15.93,170.8)(15.64,170.66)(15.34,170.51)(15.05,170.33)(14.76,170.19)(14.47,170.05)(14.22,169.89)
\path(14.22,169.89)(13.93,169.75)(13.68,169.58)(13.43,169.44)(13.21,169.3)(12.97,169.14)(12.75,169.0)(12.52,168.83)(12.31,168.69)(12.1,168.54)
\path(12.1,168.54)(11.89,168.38)(11.69,168.23)(11.51,168.08)(11.31,167.94)(11.14,167.79)(10.97,167.63)(10.8,167.48)(10.64,167.33)(10.5,167.16)
\path(10.5,167.16)(10.34,167.01)(10.19,166.86)(10.06,166.69)(9.92,166.55)(9.8,166.38)(9.67,166.23)(9.56,166.05)(9.44,165.89)(9.34,165.73)
\path(9.34,165.73)(9.22,165.55)(9.14,165.39)(9.05,165.23)(8.96,165.05)(8.88,164.86)(8.8,164.69)(8.71,164.51)(8.64,164.33)(8.59,164.16)
\path(8.59,164.16)(8.51,163.98)(8.46,163.8)(8.39,163.61)(8.36,163.41)(8.3,163.23)(8.26,163.04)(8.22,162.83)(8.19,162.63)(8.14,162.42)
\path(8.14,162.42)(8.13,162.23)(8.09,162.01)(8.06,161.8)(8.05,161.58)(8.03,161.36)(8.03,161.14)(8.01,160.92)(8.01,160.69)(8.0,160.47)
\path(8.0,160.47)(8.0,160.23)(8.0,160.0)(8.0,160.0)
\path(34.0,78.0)(34.0,78.0)(36.02,78.52)(38.0,79.04)(39.93,79.55)(41.84,80.05)(43.72,80.55)(45.56,81.04)(47.36,81.52)(49.11,81.99)
\path(49.11,81.99)(50.84,82.43)(52.54,82.88)(54.18,83.33)(55.81,83.75)(57.4,84.18)(58.95,84.58)(60.47,84.99)(61.97,85.38)(63.43,85.77)
\path(63.43,85.77)(64.86,86.15)(66.25,86.54)(67.61,86.9)(68.94,87.27)(70.23,87.61)(71.51,87.97)(72.75,88.3)(73.94,88.65)(75.15,88.97)
\path(75.15,88.97)(76.3,89.3)(77.41,89.63)(78.51,89.93)(79.58,90.25)(80.62,90.55)(81.65,90.86)(82.62,91.15)(83.58,91.44)(84.51,91.74)
\path(84.51,91.74)(85.44,92.02)(86.33,92.29)(87.19,92.58)(88.01,92.86)(88.83,93.13)(89.62,93.4)(90.4,93.66)(91.12,93.93)(91.87,94.19)
\path(91.87,94.19)(92.55,94.44)(93.23,94.72)(93.9,94.97)(94.51,95.22)(95.15,95.49)(95.73,95.74)(96.3,96.0)(96.87,96.25)(97.41,96.5)
\path(97.41,96.5)(97.93,96.75)(98.43,97.02)(98.91,97.27)(99.37,97.52)(99.8,97.77)(100.25,98.04)(100.66,98.29)(101.05,98.55)(101.44,98.8)
\path(101.44,98.8)(101.8,99.08)(102.16,99.33)(102.5,99.61)(102.8,99.88)(103.12,100.15)(103.43,100.43)(103.69,100.69)(103.98,101.0)(104.23,101.27)
\path(104.23,101.27)(104.48,101.55)(104.72,101.86)(104.94,102.15)(105.15,102.44)(105.33,102.77)(105.54,103.08)(105.72,103.4)(105.9,103.72)(106.05,104.04)
\path(106.05,104.04)(106.19,104.36)(106.36,104.69)(106.5,105.05)(106.62,105.4)(106.75,105.75)(106.87,106.11)(106.98,106.49)(107.08,106.86)(107.18,107.25)
\path(107.18,107.25)(107.26,107.63)(107.36,108.04)(107.44,108.43)(107.51,108.86)(107.58,109.27)(107.66,109.69)(107.73,110.15)(107.8,110.58)(107.87,111.05)
\path(107.87,111.05)(107.93,111.51)(108.0,111.98)(108.0,112.0)
\path(34.0,78.0)(34.0,78.0)(33.34,78.22)(32.7,78.47)(32.05,78.68)(31.45,78.91)(30.84,79.13)(30.25,79.33)(29.64,79.55)(29.09,79.77)
\path(29.09,79.77)(28.53,79.97)(27.96,80.16)(27.43,80.36)(26.88,80.55)(26.37,80.75)(25.87,80.93)(25.37,81.11)(24.87,81.29)(24.39,81.47)
\path(24.39,81.47)(23.94,81.63)(23.46,81.8)(23.03,81.97)(22.6,82.13)(22.17,82.3)(21.75,82.44)(21.31,82.61)(20.93,82.77)(20.54,82.93)
\path(20.54,82.93)(20.14,83.08)(19.77,83.22)(19.39,83.38)(19.06,83.52)(18.7,83.65)(18.35,83.8)(18.02,83.94)(17.68,84.08)(17.38,84.22)
\path(17.38,84.22)(17.06,84.36)(16.77,84.5)(16.47,84.63)(16.18,84.77)(15.92,84.9)(15.64,85.04)(15.38,85.16)(15.1,85.3)(14.85,85.43)
\path(14.85,85.43)(14.6,85.58)(14.38,85.69)(14.14,85.83)(13.93,85.97)(13.71,86.11)(13.5,86.24)(13.27,86.38)(13.09,86.52)(12.89,86.65)
\path(12.89,86.65)(12.68,86.79)(12.52,86.93)(12.34,87.05)(12.17,87.19)(12.0,87.33)(11.81,87.5)(11.67,87.63)(11.51,87.79)(11.36,87.93)
\path(11.36,87.93)(11.22,88.08)(11.09,88.24)(10.94,88.4)(10.81,88.55)(10.69,88.72)(10.56,88.88)(10.44,89.04)(10.34,89.19)(10.21,89.38)
\path(10.21,89.38)(10.11,89.54)(10.01,89.72)(9.89,89.9)(9.8,90.08)(9.71,90.27)(9.61,90.44)(9.53,90.65)(9.44,90.83)(9.36,91.04)
\path(9.36,91.04)(9.26,91.25)(9.19,91.44)(9.11,91.66)(9.03,91.88)(8.96,92.11)(8.88,92.33)(8.81,92.55)(8.75,92.79)(8.67,93.02)
\path(8.67,93.02)(8.61,93.27)(8.55,93.52)(8.47,93.77)(8.42,94.04)(8.36,94.29)(8.3,94.55)(8.22,94.83)(8.17,95.11)(8.11,95.4)
\path(8.11,95.4)(8.05,95.69)(8.0,95.99)(8.0,96.0)
\drawdot{10.0}{158.0}
\drawdot{12.0}{154.0}
\drawdot{16.0}{150.0}
\drawdot{20.0}{146.0}
\drawdot{26.0}{142.0}
\drawdot{50.0}{132.0}
\drawdot{34.0}{138.0}
\drawdot{50.0}{132.0}
\drawdot{42.0}{134.0}
\drawdot{42.0}{134.0}
\drawpath{104.0}{150.0}{104.0}{100.0}
\drawpath{52.0}{172.0}{52.0}{84.0}
\drawdot{66.0}{128.0}
\drawdot{74.0}{126.0}
\drawdot{82.0}{124.0}
\drawdot{90.0}{122.0}
\drawpath{100.0}{154.0}{100.0}{98.0}
\drawdot{10.0}{100.0}
\drawdot{96.0}{120.0}
\drawdot{12.0}{104.0}
\drawdot{106.0}{114.0}
\drawdot{20.0}{112.0}
\drawdot{16.0}{108.0}
\drawdot{74.0}{128.0}
\drawdot{102.0}{118.0}
\drawpath{94.0}{158.0}{94.0}{96.0}
\drawpath{88.0}{160.0}{88.0}{94.0}
\drawpath{82.0}{162.0}{82.0}{92.0}
\drawpath{76.0}{164.0}{76.0}{90.0}
\drawdot{82.0}{130.0}
\drawdot{26.0}{116.0}
\drawdot{90.0}{132.0}
\drawdot{34.0}{120.0}
\drawdot{106.0}{140.0}
\drawdot{102.0}{136.0}
\drawdot{96.0}{134.0}
\drawdot{42.0}{122.0}
\drawdot{50.0}{124.0}
\drawdot{58.0}{126.0}
\drawpath{70.0}{166.0}{70.0}{88.0}
\drawpath{64.0}{168.0}{64.0}{86.0}
\drawpath{58.0}{170.0}{58.0}{86.0}
\drawpath{46.0}{174.0}{46.0}{82.0}
\drawpath{40.0}{176.0}{40.0}{80.0}
\drawpath{34.0}{178.0}{34.0}{78.0}
\drawpath{28.0}{176.0}{28.0}{80.0}
\drawpath{22.0}{174.0}{22.0}{82.0}
\drawpath{16.0}{170.0}{16.0}{86.0}
\drawpath{12.0}{168.0}{12.0}{88.0}
\drawvector{56.0}{202.0}{18.0}{0}{1}
\drawvector{108.0}{60.0}{20.0}{1}{0}
\drawvector{38.0}{50.0}{10.0}{-1}{-1}
\drawdot{58.0}{130.0}
\drawcenteredtext{44.0}{220.0}{t}
\drawcenteredtext{18.0}{38.0}{$e^{1}$}
\drawcenteredtext{124.0}{52.0}{$e^{d}$}
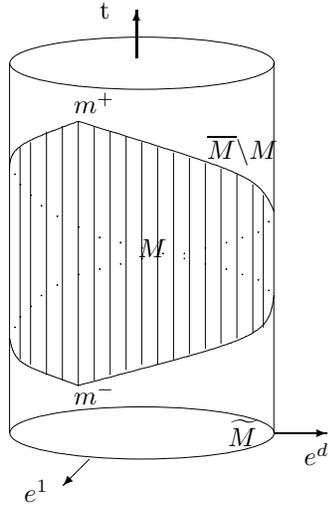
\drawcenteredtext{58.0}{6.0}{Fig.1   An embedding of the Minkowski space into the manifold $M$}
\drawcenteredtext{62.0}{130.0}{$M$}
\drawcenteredtext{96.0}{60.0}{$\widetilde {M}$}
\drawcenteredtext{40.0}{185.0}{$m^{+}$}
\drawcenteredtext{40.0}{75.0}{$m^{-}$}
\drawcenteredtext{96.0}{167.0}{$\overline {M}\backslash M$}
\end{picture}

\bigskip

\begin{picture}(160,282)
\thinlines
\drawpath{10.0}{172.0}{80.0}{242.0}
\drawpath{80.0}{242.0}{150.0}{172.0}
\drawpath{150.0}{172.0}{80.0}{102.0}
\drawpath{80.0}{102.0}{10.0}{172.0}
\drawpath{62.0}{186.0}{10.0}{134.0}
\drawpath{62.0}{150.0}{10.0}{202.0}
\drawpath{62.0}{186.0}{132.0}{116.0}
\drawpath{150.0}{202.0}{130.0}{222.0}
\drawpath{10.0}{278.0}{10.0}{58.0}
\drawpath{150.0}{278.0}{150.0}{58.0}
\drawdot{62.0}{184.0}
\drawdot{62.0}{180.0}
\drawdot{62.0}{176.0}
\drawdot{62.0}{172.0}
\drawdot{62.0}{168.0}
\drawdot{62.0}{164.0}
\drawdot{62.0}{160.0}
\drawdot{62.0}{156.0}
\drawdot{62.0}{152.0}
\drawdot{66.0}{178.0}
\drawdot{58.0}{178.0}
\drawdot{58.0}{174.0}
\drawdot{66.0}{174.0}
\drawdot{58.0}{170.0}
\drawdot{66.0}{170.0}
\drawdot{58.0}{166.0}
\drawdot{66.0}{166.0}
\drawdot{58.0}{162.0}
\drawdot{66.0}{162.0}
\drawdot{58.0}{158.0}
\drawdot{66.0}{158.0}
\drawdot{54.0}{160.0}
\drawdot{54.0}{164.0}
\drawdot{54.0}{168.0}
\drawdot{54.0}{172.0}
\drawdot{54.0}{176.0}
\drawdot{70.0}{176.0}
\drawdot{50.0}{170.0}
\drawdot{50.0}{166.0}
\drawdot{46.0}{168.0}
\drawdot{70.0}{172.0}
\drawdot{70.0}{168.0}
\drawdot{70.0}{164.0}
\drawdot{74.0}{166.0}
\drawdot{74.0}{170.0}
\drawpath{130.0}{222.0}{62.0}{150.0}
\drawpath{150.0}{134.0}{132.0}{116.0}
\drawpath{76.0}{172.0}{76.0}{172.0}
\drawdotline{10.0}{198.0}{14.0}{198.0}
\drawdotline{10.0}{194.0}{18.0}{194.0}
\drawdotline{10.0}{190.0}{22.0}{190.0}
\drawdotline{10.0}{186.0}{26.0}{186.0}
\drawdotline{10.0}{182.0}{30.0}{182.0}
\drawdotline{10.0}{178.0}{34.0}{178.0}
\drawdotline{10.0}{174.0}{38.0}{174.0}
\drawdotline{10.0}{170.0}{42.0}{170.0}
\drawdotline{10.0}{166.0}{42.0}{166.0}
\drawdotline{10.0}{162.0}{38.0}{162.0}
\drawdotline{10.0}{158.0}{34.0}{158.0}
\drawdotline{10.0}{154.0}{30.0}{154.0}
\drawdotline{10.0}{150.0}{26.0}{150.0}
\drawdotline{10.0}{146.0}{22.0}{146.0}
\drawdotline{10.0}{142.0}{18.0}{142.0}
\drawdotline{10.0}{138.0}{14.0}{138.0}
\drawdotline{128.0}{220.0}{132.0}{220.0}
\drawdotline{124.0}{216.0}{136.0}{216.0}
\drawdotline{120.0}{212.0}{140.0}{212.0}
\drawdotline{116.0}{208.0}{144.0}{208.0}
\drawdotline{114.0}{204.0}{148.0}{204.0}
\drawdotline{110.0}{200.0}{150.0}{200.0}
\drawdotline{106.0}{196.0}{150.0}{196.0}
\drawdotline{102.0}{192.0}{150.0}{192.0}
\drawdotline{98.0}{188.0}{150.0}{188.0}
\drawdotline{94.0}{184.0}{150.0}{184.0}
\drawdotline{90.0}{180.0}{150.0}{180.0}
\drawdotline{86.0}{176.0}{150.0}{176.0}
\drawdotline{84.0}{172.0}{150.0}{172.0}
\drawdotline{82.0}{168.0}{150.0}{168.0}
\drawdotline{84.0}{164.0}{150.0}{164.0}
\drawdotline{88.0}{160.0}{150.0}{160.0}
\drawdotline{92.0}{156.0}{150.0}{156.0}
\drawdotline{96.0}{152.0}{150.0}{152.0}
\drawdotline{100.0}{148.0}{150.0}{148.0}
\drawdotline{104.0}{144.0}{150.0}{144.0}
\drawdotline{108.0}{140.0}{150.0}{140.0}
\drawdotline{112.0}{136.0}{150.0}{136.0}
\drawdotline{116.0}{132.0}{148.0}{132.0}
\drawdotline{120.0}{128.0}{144.0}{128.0}
\drawdotline{124.0}{124.0}{140.0}{124.0}
\drawdotline{128.0}{120.0}{136.0}{120.0}
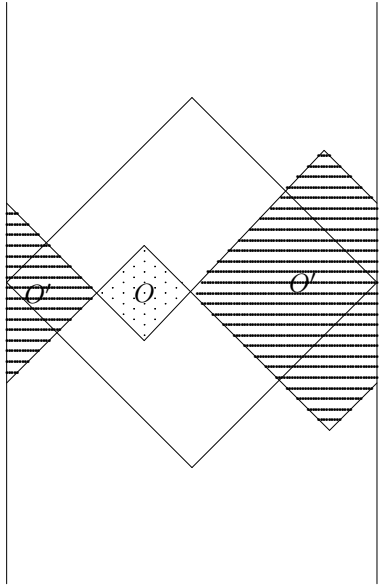
\drawcenteredtext{80.0}{6.0}{Fig.2   The spacelike complement of a double cone $O$ in $M$ within $\widetilde {M}$}
\drawcenteredtext{62.0}{168.0}{$O$}
\drawcenteredtext{22.0}{168.0}{$O^{\prime}$}
\drawcenteredtext{122.0}{172.0}{$O^{\prime}$}
\end{picture}

\bigskip

On the other hand the living space of the observable algebra is the
Dirac-Weyl compactification $\bar{M}$ of $M$ which is depicted as Fig.3 with
opposite two sides $a$ and $b$ identified. Vice versa the Minkowski space
results from puncturing the Dirac-Weyl compactification $\bar{M}$ at $%
m_{+}=m_{-}$ and simultaneously removing the whole subtended lightlike d-1
dimensional subspace. Note that as a result of the identification the union
of the timelike and spacelike complements form a connected set in $\bar{M}$;
in fact that part of the spacelike complement of a double cone $\mathcal{O}$
in the covering $\widetilde{M}$ beyond $M$ becomes converted into the
Huygens timelike region with respect to $\mathcal{O}.$ The first use of
these geometrical properties in the setting of algebraic QFT is due to
Hislop and Longo \cite{HL}. The pictures are closely related to those used
by Penrose, except that Penrose does not use them for compactification since
he is dealing with a conformal class of spacetime metrics and not with
conformal invariant observable matter fulfilling the Huygens principle.

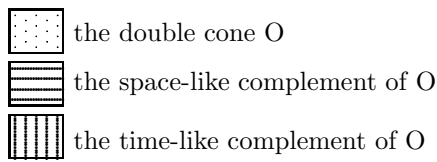
\begin{picture}(176,266)
\thinlines
\drawpath{8.0}{178.0}{88.0}{258.0}
\drawpath{88.0}{258.0}{168.0}{178.0}
\drawpath{168.0}{178.0}{88.0}{98.0}
\drawpath{88.0}{98.0}{8.0}{178.0}
\drawpath{18.0}{168.0}{98.0}{248.0}
\drawpath{24.0}{194.0}{104.0}{114.0}
\drawpath{118.0}{228.0}{38.0}{148.0}
\drawpath{48.0}{218.0}{128.0}{138.0}
\drawdot{58.0}{206.0}
\drawdot{58.0}{202.0}
\drawdot{58.0}{198.0}
\drawdot{58.0}{194.0}
\drawdot{58.0}{190.0}
\drawdot{58.0}{186.0}
\drawdot{58.0}{182.0}
\drawdot{58.0}{178.0}
\drawdot{58.0}{174.0}
\drawdot{58.0}{170.0}
\drawdot{54.0}{200.0}
\drawdot{62.0}{200.0}
\drawdot{54.0}{196.0}
\drawdot{54.0}{192.0}
\drawdot{54.0}{188.0}
\drawdot{54.0}{184.0}
\drawdot{54.0}{180.0}
\drawdot{54.0}{176.0}
\drawdot{54.0}{172.0}
\drawdot{54.0}{168.0}
\drawdot{62.0}{196.0}
\drawdot{62.0}{192.0}
\drawdot{62.0}{188.0}
\drawdot{62.0}{184.0}
\drawdot{62.0}{180.0}
\drawdot{62.0}{176.0}
\drawdot{66.0}{198.0}
\drawdot{66.0}{194.0}
\drawdot{66.0}{190.0}
\drawdot{66.0}{186.0}
\drawdot{66.0}{182.0}
\drawdot{66.0}{178.0}
\drawdot{70.0}{184.0}
\drawdot{70.0}{188.0}
\drawdot{70.0}{192.0}
\drawdot{74.0}{190.0}
\drawdot{74.0}{186.0}
\drawdot{50.0}{198.0}
\drawdot{50.0}{194.0}
\drawdot{50.0}{190.0}
\drawdot{50.0}{186.0}
\drawdot{50.0}{182.0}
\drawdot{50.0}{178.0}
\drawdot{50.0}{174.0}
\drawdot{50.0}{170.0}
\drawdot{46.0}{192.0}
\drawdot{46.0}{188.0}
\drawdot{46.0}{184.0}
\drawdot{46.0}{180.0}
\drawdot{46.0}{176.0}
\drawdot{42.0}{178.0}
\drawdot{42.0}{182.0}
\drawdot{42.0}{186.0}
\drawdot{42.0}{190.0}
\drawdot{38.0}{184.0}
\drawdotline{118.0}{226.0}{118.0}{226.0}
\drawdotline{118.0}{226.0}{120.0}{226.0}
\drawdotline{114.0}{222.0}{122.0}{222.0}
\drawdotline{108.0}{218.0}{128.0}{218.0}
\drawdotline{104.0}{214.0}{132.0}{214.0}
\drawdotline{100.0}{210.0}{136.0}{210.0}
\drawdotline{96.0}{206.0}{140.0}{206.0}
\drawdotline{92.0}{202.0}{144.0}{202.0}
\drawdotline{88.0}{198.0}{148.0}{198.0}
\drawdotline{84.0}{194.0}{152.0}{194.0}
\drawdotline{80.0}{190.0}{156.0}{190.0}
\drawdotline{80.0}{186.0}{160.0}{186.0}
\drawdotline{84.0}{182.0}{164.0}{182.0}
\drawdotline{88.0}{178.0}{168.0}{178.0}
\drawdotline{92.0}{174.0}{164.0}{174.0}
\drawdotline{96.0}{170.0}{160.0}{170.0}
\drawdotline{100.0}{166.0}{156.0}{166.0}
\drawdotline{104.0}{162.0}{152.0}{162.0}
\drawdotline{108.0}{158.0}{148.0}{158.0}
\drawdotline{112.0}{154.0}{144.0}{154.0}
\drawdotline{116.0}{150.0}{140.0}{150.0}
\drawdotline{120.0}{146.0}{136.0}{146.0}
\drawdotline{124.0}{142.0}{132.0}{142.0}
\drawdotline{22.0}{192.0}{26.0}{192.0}
\drawdotline{18.0}{188.0}{30.0}{188.0}
\drawdotline{14.0}{184.0}{34.0}{184.0}
\drawdotline{10.0}{180.0}{30.0}{180.0}
\drawdotline{10.0}{176.0}{26.0}{176.0}
\drawdotline{14.0}{172.0}{22.0}{172.0}
\drawdotline{50.0}{220.0}{50.0}{216.0}
\drawdotline{54.0}{224.0}{54.0}{212.0}
\drawdotline{58.0}{228.0}{58.0}{208.0}
\drawdotline{62.0}{232.0}{62.0}{212.0}
\drawdotline{66.0}{236.0}{66.0}{216.0}
\drawdotline{70.0}{240.0}{70.0}{220.0}
\drawdotline{74.0}{244.0}{74.0}{224.0}
\drawdotline{78.0}{248.0}{78.0}{228.0}
\drawdotline{82.0}{252.0}{82.0}{232.0}
\drawdotline{86.0}{256.0}{86.0}{236.0}
\drawdotline{90.0}{256.0}{90.0}{240.0}
\drawdotline{94.0}{252.0}{94.0}{244.0}
\drawdotline{40.0}{150.0}{40.0}{146.0}
\drawdotline{44.0}{154.0}{44.0}{142.0}
\drawdotline{48.0}{158.0}{48.0}{138.0}
\drawdotline{52.0}{160.0}{52.0}{134.0}
\drawdotline{56.0}{162.0}{56.0}{130.0}
\drawdotline{60.0}{158.0}{60.0}{126.0}
\drawdotline{64.0}{154.0}{64.0}{122.0}
\drawdotline{68.0}{150.0}{68.0}{118.0}
\drawdotline{72.0}{146.0}{72.0}{114.0}
\drawdotline{76.0}{142.0}{76.0}{110.0}
\drawdotline{80.0}{138.0}{80.0}{106.0}
\drawdotline{84.0}{134.0}{84.0}{102.0}
\drawdotline{88.0}{130.0}{88.0}{98.0}
\drawdotline{92.0}{126.0}{92.0}{102.0}
\drawdotline{96.0}{122.0}{96.0}{106.0}
\drawdotline{100.0}{118.0}{100.0}{110.0}
\drawvector{92.0}{94.0}{80.0}{1}{1}
\drawvector{84.0}{94.0}{80.0}{-1}{1}
\drawvector{4.0}{182.0}{78.0}{1}{1}
\drawvector{172.0}{182.0}{80.0}{-1}{1}
\drawcenteredtext{28.0}{236.0}{a}
\drawcenteredtext{150.0}{118.0}{a}
\drawcenteredtext{150.0}{240.0}{b}
\drawcenteredtext{22.0}{120.0}{b}
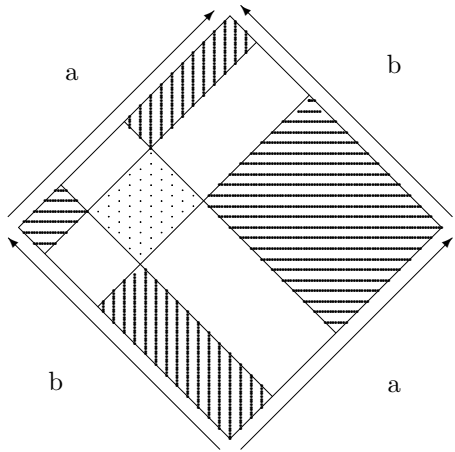
\drawcenteredtext{88.0}{72.0}{Fig.3   The Dirac-Weyl compactification}
\drawlefttext{54.0}{52.0}{the double cone O}
\drawlefttext{54.0}{32.0}{the space-like complement of O}
\drawlefttext{54.0}{10.0}{the time-like complement of O}
\drawframebox{40.0}{52.0}{20.0}{16.0}{}
\drawframebox{40.0}{32.0}{20.0}{16.0}{}
\drawframebox{40.0}{12.0}{20.0}{16.0}{}
\drawdot{32.0}{58.0}
\drawdot{32.0}{54.0}
\drawdot{32.0}{50.0}
\drawdot{32.0}{46.0}
\drawdot{36.0}{48.0}
\drawdot{36.0}{52.0}
\drawdot{36.0}{56.0}
\drawdot{40.0}{58.0}
\drawdot{40.0}{54.0}
\drawdot{40.0}{50.0}
\drawdot{40.0}{46.0}
\drawdot{44.0}{48.0}
\drawdot{44.0}{52.0}
\drawdot{44.0}{56.0}
\drawdot{48.0}{58.0}
\drawdot{48.0}{54.0}
\drawdot{48.0}{50.0}
\drawdot{48.0}{46.0}
\drawdotline{30.0}{38.0}{50.0}{38.0}
\drawdotline{30.0}{34.0}{50.0}{34.0}
\drawdotline{30.0}{30.0}{50.0}{30.0}
\drawdotline{30.0}{26.0}{50.0}{26.0}
\drawdotline{32.0}{20.0}{32.0}{4.0}
\drawdotline{36.0}{20.0}{36.0}{4.0}
\drawdotline{40.0}{20.0}{40.0}{4.0}
\drawdotline{44.0}{20.0}{44.0}{4.0}
\drawdotline{48.0}{20.0}{48.0}{4.0}
\end{picture}

It is hard to resist mentioning that the d+2 dimensional setting for the
compactification and subsequent covering of d-dimensional Minkowski
spacetime also lends itself to obtain a natural relation with the dd+1
dimensional $AdS$ (anti de Sitter) spacetime by taking instead of the
surface of the light cone a hyperbolic region inside this light cone $\xi
^{2}=1.$ It is immediately clear that asymptotically this $\left( D+1\right) 
$ dimensional Lorentzian noncompact manifold in the associated $\xi $%
-parametrization has a d-dimensional conformal boundary which in the above
picture corresponds to the asymptotic coalescence of the $AdS_{D+1}$ with
the light cone directions. This asymptotic pointlike relation can of course
not be continued inside the $AdS_{D+1}$ spacetime, but the action of the
same $SO(D,2)$ group on the two manifolds suggests a relation between
d-dimensional conformal double cone (conformal transforms of Rindler wedge
regions) and d+1 dimensional wedge regions on $AdS_{D+1}$ i.e. between those
regions which result from projecting d+2 dimensional wedges in the ambient
space (on which there is a transitive linear action of $SO(D,2)$) onto $\bar{%
M}_{D}$ and $AdS_{D+1}.$Using the setting of algebraic QFT, Rehren \cite{Re1}
converted this geometric relation into an isomorphism between algebraic
quantum field theories where the principle objects are localized algebras
and not their coordinatizations in terms of pointlike field generators. In
fact as one naively expects the isomorphism negates a relation between
pointlike field theories (see below).

Although from a logical point of view the above observation on AdS spacetime
belongs naturally to the conformal compactification setting of the 70s where
most of the above observations were made, history (as almost always and in
particular in this case) did not follow logic. Rather part of the
isomorphism namely the mapping $AdS_{D+1}\overset{asympt.}{\rightarrow }\bar{%
M}_{D}$ for the corresponding QFTs was first observed by string theorist at
the end of the 90's \cite{Wit} in connection with their speculative ideas on
quantum gravity. Although the underlying idea was that the asymptotic map
characterizes a unique AdS theory once the conformal asymptote has been
prescribed, it was Rehren's construction which supplied a constructive
mathematical proof for the full isomorphism and at the same time also
highlighted the conceptual scope of (field-coordinate-free) local quantum
physics.

The isomorphism can be brought closer to the realm of particle physics if
one highlights the above analogy between the Hamiltonian $H$ and the
conformal ``Hamiltonian'' $R_{0}$ by asking the following question: is there
a QFT which maintains the symmetries but for which the conformal Hamiltonian
of the $M_{D}$ becomes the true Hamiltonian? The answer is unique, it is
precisely the same $AdS_{D+1}$ theory of the Rehren isomorphism which
changes the physical interpretation and the spacetime affiliation, but not
the group theoretical and algebraic net structure.

The particle physics nature of this isomorphism can be further clarified by
studying concrete examples e.g. the fate of free conformal/AdS fields under
this isomorphism. The result is rather interesting \cite{Notes}. A pointlike
AdS free fields has too many degrees of freedom with the result that its
conformal image is a special ``generalized free field'' with a homogenous
mass distribution which destroys the primitive causality i.e. the
requirement that the algebraic data in a time slice (in order to avoid short
distance problems of spacelike surfaces) which covers a compact spatial
region fix the data in the ``causal shadow'' region (the double cone shaped
causal envelop). Using Rehren's graphical illustration \cite{Re2}which
depicts conformal QFT on a cylindrical boundary of the AdS world, one sees
that there are more and more degrees of freedom from the inside of AdS
entering the causal shadow as one moves upward in time. Vice versa if one
starts from a free conformal theory than the AdS image can not sustain
pointlike fields but only configurations which are delocalized (and hence
diluted) in one direction as some kind of Nielsen-Olsen string i.e. the
memory from its d-dimensional pointlike origin is stored in the AdS image.
This means in practical terms that the nice idea to start with a Lagrangian
AdS theory end re-process them with the Witten prescription to the conformal
side in order to enrich the set of conformal models does not work. One
obtains conformal theories in this way but they are unphysical, a fact which
remains concealed in the euclidean formulation.

This also affects the conjectures string theory--SYM relation to the extend
that it relies on a relation between $AdS_{5}$ and conformal SYM Lagrangian
theories.

The algebraic isomorphism itself is not affected limited by these
deficiencies in pointlike (Lagrangian) relations and their remains the
intellectually challenging problem of understanding the algebraic conformal
decomposition theory and its DHR origin directly for QFT on AdS. Just
because the isomorphism changes the interpretation, it is by no means
obvious without looking carefully at details, what the observations in this
article mean in a AdS spacetime. Since the spectrum of anomalous dimensions
becomes encoded into the spectrum of a Hamiltonian one even could be
optimistic and hope for certain simplifications on the AdS side. A rather
trivial and unfortunately atypical case is $AdS_{2}$ whose bona fide
Hamiltonian is the Virasoro $L_{0}$ and in fact it is the only system with a
maximal SL(2,R) symmetry having that compact (discrete spectrum) opaerator
as its hamiltonian.

The above analogies between the d=1+1 case and the higher dimensional
conformal field theory should however not lead one into overlooking a
remarkable difference. Already on a purely classical level the
characteristic value problem for the free wave equation is totally different
from either its massive counterpart or from the d\TEXTsymbol{>}2 case.
Whereas in the latter cases the data on one lightray or lightfront is
complete, the zero mass d=1+1 case needs both the lightray data in order to
determine the d=1+1 theory. In the QFT the manifestation of this is the
tensor factorization into the chiral degrees of freedom which amounts to a
doubling of degrees of freedom. In the next section we will see that this
also leads to an exceptional behavior in the timelike Huygens structure and
the associated timelike braid group structure. So the chirally factorizing
d=1+1 situation is a guide in certain higher dimensional aspects and stands
in interesting contrast to others.

\section{Central decomposition and braid group structure}

In the previous section we have emphasized certain analogies between the
timelike algebraic structure in higher dimensional and chiral conformal
theories. For the latter the decomposition theory has a more fundamental
explanation in terms of a local plektonic superselection structure. This in
particular means that the components appearing in (\ref{comp}) permit a
local refinement; they can be further reduces into DHR localized charge
sectors of an observable local algebra which lives on $S^{1}$ and which in
concrete models is generated by the energy momentum tensor, current
algebras, W-algebras etc. The commutation structure of charge-carrying
chiral fields obtained by the DHR method is an exchange algebra

\begin{eqnarray}
F_{\alpha ,\beta }(x)G_{\beta ,\gamma }(y) &=&\sum_{\beta ^{\prime
}}R_{\beta ,\beta ^{\prime }}^{(\alpha ,\gamma )}[c(F),c(G)]G_{\alpha ,\beta
^{\prime }}(y)F_{\beta ^{\prime },\gamma }(x),\,\,x>y  \label{time} \\
F_{a,\beta }G_{\beta ,\gamma } &=&\sum_{\beta ^{\prime }}R_{\beta ,\beta
^{\prime }}^{(\alpha ,\gamma )}[c(F),c(G)]G_{\alpha ,\beta ^{\prime
}}F_{\beta ^{\prime },\gamma },\,\,locF>locG
\end{eqnarray}
where in the second line we have used the more general operator formulation
of AQFT and the c's in the bracket $R[c(F),c(G)]$ denotes the dependence of
the R-matrices on the superselected charges c of the participating
operators. The localization is always meant relative to the observables \cite
{FRSII}. The Artin relations\footnote{%
In the physical literature they are often called Yang-Baxter relations, but
our conceptual fidelity prohibits us to verbally mix up the S-matrix
Yang-Baxter setting with the plektonic statistics concepts within Artin's
braid group.} are a consequence of the associativity of this algebra. Here
the indices $\alpha ,\beta ,\gamma $ refers to projectors on irreducible DHR
representation spaces. Although the latter are a refinement of the
projectors appearing in the center $Z$, we will for simplicity in writing
maintain the same notation.

The validity of such an exchange algebra structure would supply a natural
local explanation for the center superselection rules. We will therefore
postulate this structure for the timelike region in higher dimensional
conformal theories (where now $\lessgtr $ to $\mp $timelike ordering) and,
in lack of concrete examples, test its consistency. It is reasonable to
start with consistency checks in the standard Wightman framework of
pointlike covariant fields.

The most powerful tool of Wightman's formulation is provided by the analytic
properties of correlation functions. It is well known that the complexified
Lorentz group may be used to extend the tube analyticity associated with the
physical positive energy-momentum spectrum. The famous BHW theorem \cite
{St-Wi} insures that this extension remains univalued in a new complex
domain and the Jost theorem characterizes its real points. Finally spacelike
locality links the various permutations of the position field operators
within the correlation function to one permutation (anti)symmetric analytic
master function which is still univalued. The various correlation functions
on the physical boundary with different operator ordering can be obtained by
different temporal $i\varepsilon $ prescriptions descending to the real
boundary from within the tube.

Complexifying the scale transformations, the conformal correlations can be
extended into a still bigger analyticity region which even incorporates
``timelike Jost points''. But trying to find a univalued master function
which links the various orders together fails in the presence of fields with
anomalous dimensions and only works for observable fields which are local
fields on the compactification $\bar{M}.$ The latter are the analogs of
chiral observables, except that apart from (composite) free fields one does
not have algebraic examples since Virasoro- and Kac-Moody algebras do not
exist in higher dimensions. The timelike braid group structure suggest that
the role of the permutation group in the analytic extension from timelike
points should be replaced by the braid group. The resulting ramification
takes place in a region which is obtained analytic extension by the complex
dilation group, which leaves the old univalued BHW region free of branch
cuts. Therefore the timelike braid structure is consistent with the BHW
analyticity structure. Let us look at other arguments which test the
consistency of the old spacelike locality with the new timelike localization
structure because the problem of coexistence of these two regions is the
main difference to the chiral case. This significant difference even remains
if one glues together the two chiralities to a 2-dimensional conformal
theory. As already mentioned before, it has its origin already in the
classical wave propagation theory one needs two sets of characteristic light
ray data to e.g. determine the amplitudes in a wedge region whereas for any
higher dimensional theory (and even massive d=1+1 propagation) one light
front is enough.

A plektonic charge structure which is only visible in the timelike region
would immediately explain the appearance of a nontrivial timelike center and
the spectrum of anomalous dimension. It would sort of ``kinematize''
conformal interactions and reveal conformal QFTs as basically free theories
if it would not be for that part of interaction which sustains the timelike
plektonic structure. Of course the situation trivializes if the theory has
no anomalous dimensions and nontrivial components. Analogous to \cite{Haag}
(remarks at end of section V.4) we conjecture that this characterizes
interaction free conformal theories which are generated by free fields%
\footnote{%
Note that this conjecture would be wrong in D=1+1 since from selfdual
lattice construction on current algebras one obtains models without
nontrivial sectors which are different from free fields.}. What makes this
issue somewhat complicated is the fact that contrary to chiral theories we
do not have a single nontrivial example because this issue is neither
approachable from the representation theory of known infinite dimensional
Lie-algebras nor from the formal euclidean functional integral method. The
remaining strategy is to show structural consistency of the spacelike local-
with the conjectured timelike plektonic- structure and to find a new
construction method (non energy-momentum tensor- or current- algebra based,
non-Lagrangian). Here we are mainly be concerned with consistency arguments
and in the following we will comment how local/plektonic on-vacuum relations
between two fields can be commuted through to a generic position.

Assume for simplicity, as we have already tacitly done in our decomposition
formulas before, that we are in a ``minimalistic'' situation (similar to
minimal models or W-algebras in the chiral setting) where the field theory
has no internal symmetry group. Actually the whole discussion can be carried
out in the presence of nontrivial inner symmetries\footnote{%
I am indepted to Karl-Henning Rehren for showing me how to extend the
timelike R-matrix formalism in the presence of inner symmetry groups.}, but
the additional complications do not essentially alter the following
consistency considerations. So we assume that the fields can be given
``timelike'' charge indices $\alpha ,\beta ,\gamma ..$ and their conjugates $%
\bar{\alpha},\bar{\beta},\bar{\gamma}....$ resulting from projectors on
charge spaces so that the decomposition is as in the chiral case where the
charge projectors with the same phase factors $e^{2\pi i\delta }$ constitute
a refinement of a central projector. Clearly $\alpha $ and its conjugate $%
\bar{\alpha}$ contribute tu the same central projector. In fact we may take
over a substantial part of the formalism and concepts of \cite{R-S}. if one
replaces the chiral translation+dilation augmented by the circular rotation
generator $L_{0}$ by the spacetime symmetry group which leaves the timelike
infinite point fixed (Poincar\'{e}+dilations) augmented by the generator of
conformal time $R_{0}$ instead of the chiral $L_{0}.$ One would of course
also have to change the title of the old paper from ``Einstein causality and
Artin braids'' to ``Huygens causality and Artin braids'' referring to the
timelike ordering for which the conformal observables fulfill the Huygens
principle of vanishing commutators. The ``on-vacuum'' structure of
commutation relations follows from the structure of the conformal 3-point
functions 
\begin{eqnarray}
&&\left\langle H^{\ast }(x_{3})G(x_{2})F(x_{1})\right\rangle =c_{FGH}\frac{1%
}{\left[ -(x_{12})_{\varepsilon }^{2}\right] ^{\delta _{3}}}\frac{1}{\left[
-(x_{13})_{\varepsilon }^{2}\right] ^{\delta _{2}}}\frac{1}{\left[
-(x_{23})_{\varepsilon }^{2}\right] ^{\delta _{1}}} \\
&&\delta _{1}=\frac{1}{2}(\delta _{G}+\delta _{H}-\delta _{F}),\,\delta _{2}=%
\frac{1}{2}(\delta _{F}+\delta _{H}-\delta _{G}),\,\delta _{3}=\frac{1}{2}%
(\delta _{F}+\delta _{G}-\delta _{H})  \notag
\end{eqnarray}
where the $\varepsilon $-prescription was explained in the introduction. For
spacelike and timelike distances one concludes 
\begin{equation}
G(x_{2})F(x_{1})\Omega =\left\{ 
\begin{array}{c}
F(x_{1})G(x_{2})\Omega ,\,\left( x_{2}-x_{1}\right) ^{2}<0 \\ 
e^{\pi i(\delta _{F}+\delta _{G})}Z^{\ast }F(x_{1})G(x_{2})\Omega ,\,\left(
x_{2}-x_{1}\right) ^{2}>0,\,\left( x_{2}-x_{1}\right) _{0}>0
\end{array}
\right. 
\end{equation}
since this relation is valid on all quasiprimary composites $H.$ They
consist by definition of the equal point limit of the associated primary $%
H_{min}$ (lowest scale dimension operator in a superselected charge class)
multiplied with a polynomial in the observable field. These composites
applied to the vacuum form a dense set in the respective charge sector%
\footnote{%
With a bit more work and lengthier formulas one can avoid the colliding
point limit and use correlation functions containing 3 charged fields and an
arbitrary number of neutral observable fields. The dependence on the
observable coordinates is described by a rational function on $\bar{M}.$}
and hence the on-vacuum formula is a consequence of the structure of 3-point
functions. The spacelike local commutativity off-vacuum is consistent with
that on-vacuum since for $y$ timelike with respect to a spacelike pair $%
x_{1},x_{2}$ we have (here $c(\cdot )$ denote the superselected charges of
the participating operators) 
\begin{eqnarray*}
P_{\alpha }F(x_{1})G(x_{2})H(y)\Omega  &=&\sum_{\beta }P_{\alpha
}F(x_{1})P_{\beta }G(x_{2})H(y)\Omega  \\
&=&\sum_{\beta }P_{\alpha }F(x_{1})P_{\beta }e^{i\pi (\delta _{B}+\delta
_{C}-\delta _{\beta })}H(y)G(x_{2})\Omega  \\
&=&\sum_{\beta \beta ^{\prime }}R_{\beta \beta ^{\prime }}^{(\alpha \gamma
)}(c_{F},c_{G})e^{i\pi (\delta _{G}+\delta _{H}-\delta _{\beta })}P_{\alpha
}H(y)P_{\beta ^{\prime }}F(x_{1})P_{\gamma }G(x_{2})\Omega 
\end{eqnarray*}
and therefore the off-vacuum vanishing of the $F$-$G$ commutator is
consistent with the on-vacuum vanishing of this commutator if there holds a
certain relation between $R(c_{F},c_{G})$ and $R(c_{G};c_{F})$ which is
identically fulfilled for $c_{F}=c_{G}$. Similarly one does not run into
inconsistencies if one tries to obtain a timelike off-vacuum $F$-$G$
situation from the on-vacuum placement by commuting through a $H$ which is
spacelike to the timelike $F$-$G$ pair 
\begin{eqnarray}
&&P_{\alpha }F(x_{1})G(x_{2})H(y)\Omega =P_{\alpha
}H(y)F(x_{1})G(x_{2})\Omega =P_{\alpha }H(y)e^{i\pi (\delta _{B}+\delta
_{C}-\delta _{\beta })}G(x_{2})F(x_{1})\Omega   \notag \\
&=&\sum_{\beta }R_{\beta \beta ^{\prime }}P_{\alpha }G(x_{2})P_{\beta
^{\prime }}F(x_{1})H(y)\Omega =\sum_{\beta \beta ^{\prime }}R_{\beta \beta
^{\prime }}P_{\alpha }G(x_{2})P_{\beta ^{\prime }}H(y)F(x_{1})\Omega 
\end{eqnarray}
where in the second line we commuted $F$ through $G$ \textit{before} trying
to bring both to the vacuum. Since their is no rule to commute the $%
P_{\alpha }G(x_{2})P_{\beta ^{\prime }}$ with $P_{\beta ^{\prime }}H$ for $%
\left( x_{2}-y\right) ^{2}<0,$ there is no way to get to the same $HGF$
order as in the first line and hence no consistency relation is to be
checked. The absence of rules for spacelike commutations for projected
fields protects the formalism to run into inconsistencies. If components for
spacelike distances would commute than they belong to the vacuum.

Let us also briefly look at the compatibility of the timelike plektonic
structure with the conformal structure of the 4-point function of 4
identical Hermitian fields

\begin{align}
& W(x_{4},x_{3},x_{2},x_{1}):=\sum_{\gamma }\left\langle
F(x_{4})F(x_{3})P_{\gamma }F(x_{2})F(x_{1})\right\rangle \\
& =\left[ \frac{x_{42}^{2}x_{31}^{2}}{(x_{43})_{\varepsilon
}^{2}(x_{32})_{\varepsilon }^{2}(x_{21})_{\varepsilon
}^{2}(x_{14})_{\varepsilon }^{2}}\right] ^{\delta _{F}}\sum_{\gamma
}w_{\gamma }(u,v),\,\,  \notag \\
\,\,\,u& =\frac{x_{43}^{2}x_{21}^{2}}{(x_{42})_{\varepsilon
}^{2}(x_{31})_{\varepsilon }^{2}},\,\,v=\frac{x_{32}^{2}x_{41}^{2}}{%
(x_{42})_{\varepsilon }^{2}(x_{31})_{\varepsilon }^{2}}  \notag
\end{align}
Whereas the spacelike commutations leads to functional relations for $%
w=\sum_{\gamma }w_{\gamma }(u,v)$ with the exchange of two fields causing a
rational transformation of the $u,v$ (apart from multiplying the $w$ by
rational $u,v$ factors)$,$ the timelike commutation of the off-vacuum fields 
$x_{2}\leftrightarrow x_{3}$ produces rational transformation together with
R-matrix mixing of the $\gamma $-components leading to nontrivial
monodromies 
\begin{equation*}
w_{\gamma }(u,v)=\sum_{\gamma ^{\prime }}R_{\gamma \gamma ^{\prime
}}w_{\gamma ^{\prime }}(\frac{1}{u},\frac{v}{u})u^{2\delta _{F}}
\end{equation*}
As in the chiral case \cite{R-S} these relations characterize equivalence
classes of operators carrying the same superselected charges (conformal
blocks) and the selection of individual correlation functions have to be
made by using in addition their short distance behavior in terms of scaling
dimensions. Despite some similarities with the chiral case, the dependence
of $w_{\gamma }$ on two instead of one cross ratios requires the use of more
elaborate techniques (Mellin-Barnes techniques, generalized hypergeometric
functions) than the hypergeometric formalism (which is sufficient for the
chiral one variable cross ratio dependence). We find it conceivable that in
higher dimensions even for minimalistic models there could be deformation
parameters (coupling constants) which may not show up in continuously
changing anyonic phases. Here we will not pursue this matter.

Let us finally take notice of what the algebraic approach can add to these
consistency consideration.

One possible point of departure for the algebraic approach would be to start
from a Doplicher-Roberts field net $\mathcal{F}$ on $M$ and use its assumed
local conformal invariance in order to construct a unique extension $%
\widetilde{\mathcal{F}}$ on the covering $\widetilde{M}.$ This has been done
in the work \cite{BGL} where it was also shown that extended net fulfills
the important property of Haag duality which in the bosonic case is a
maximalization of causality 
\begin{equation}
\widetilde{\mathcal{F}}(\mathcal{O}^{\prime })=\widetilde{\mathcal{F}}(%
\mathcal{O})^{\prime }
\end{equation}
where we have used the standard notation of AQFT: a dash on the algebra
means the von Neumann commutant in the Hilbert space of the operator
algebra, whereas on the localization region $\mathcal{O}$ (double cone, or
any conformal transform thereof) it means spacelike complement in the global
sense of $\widetilde{M}.$ In order to obtain information about a timelike
braid group structure we need to identify an observable subnet $\mathcal{A}$
on $M$ consisting of operators which commute with $Z$ and is Haag dual for
timelike distances such that the restriction of the field net to the
observable net decomposes into irreducible representations of the latter
which obey braid group fusion laws. The projectors onto those irreducible
subspaces would than be the desired refinements of the central $Z$
projectors by which one obtains the $F_{\alpha ,\beta }$ exchange algebra
operators.

A prerequisite for this idea to work would be the existence of an autonomous
timelike Haag dual net which is sufficiently nonlocal in the spacelike sense
i.e. sufficiently different from a spacelike Haag dual net. This is
necessary in order to obtain sectors which are different from those of the
spacelike based DHR theory. A theory of observables $\mathcal{A}$ on the
compactification $\bar{M}$ is automatically Haag dual, but the localization
of the commutant of a double cone would consist of a spacelike and timelike
part \cite{Haag}. We would need a dualization which involves only the
timelike complement in $M$ of double cones and which should be sufficiently
nonlocal with respect to a conformal DHR dualization based on double cone
algebras defined in terms of intersections of infinitely many wedges.
Algebraically the Haag dual net on $\bar{M}$ results from conformal
transformations from the wedge algebra, whereas the double cones of the
timelike dual net are obtained from intersections of the forward- with the
shifted backward light cone algebra in $M.$

From a pointlike field point of view the difference in these algebras is
related to the way in which smeared fields are used to generate algebra. If
they are used on $M$ then the net at infinity is diluted because the test
functions vanish on $\bar{M}\backslash M$ and in order to re-establish the
duality balance on has to make the algebras in the net bigger which do not
contain points at infinity. This and the ensuing loss of spacelike
commutativity of these algebras has been studied in \cite{HL}. A detailed
investigation of ``cutting holes'' into the circle of chiral nets and then
producing new conformal nets from the Haag dualization of the punctured nets
can be found in \cite{GLW}. Although such studies of chiral theories cannot
answer the above consistency problems in higher dimensional theories, they
do supply valuable concepts and mathematical methods for further studies.

As in many other areas of field theory, consistency problems find their
satisfactory solutions only through mathematically controllable model
constructions and the structural analysis is only a preparatory step in the
classification and construction of models outside the standard Lagrangian
realm. Its main purpose is to prepare our intuition in an area where it is
presently underdeveloped or beset by prejudices.

Thinking of what could be the right setting for such constructions, only one
idea seems to offer sufficient conceptual depth and mathematical power. This
the modular localization approach based on the Tomita-Takesaki modular
theory and using new physical concepts as ``polarization-free generators''
of wedge algebras. There are two ways of using that theory and both are
independent of spacetime dimensions \cite{Essay}\cite{BQ}, although special
situations in low dimensions greately favor their analytical control.

One approach is based on the fact that the S-matrix is a relative modular
invariant of wedge algebras and is behind the bootstrap-formfactor
constructions in d=1+1 factorizing models. Conformal theories do not have an
S-matrix, but there are indications that their TCP operators ($\simeq $%
modular conjugations for the modular theory of the forward light cone
algebra) can be related to a simpler reference situation by generalized
``twist operators'' which contain the braid information. This idea should be
first tested for chiral theories since even there exists up to date no
systematic method to construct a model associated with a given plektonic
superselection structure. In contradistinction to the standard methods which
start from concrete observable algebras (Virasoro-, current- W-algebras) and
introduce the charge-carrying fields as intertwiners between the irreducible
representations, the modular method aims directly at the charged fields. It
is analogous to the Wigner Poincar\'{e} group representation approach to
free fields where the charged fields came first and the group invariant
observable algebras (which are more complicated since they involve composite
fields) were later introduces only for purposes of a better structural
understanding. The modular program is in a certain sense a continuation of
the Wigner approach in the presence of interactions.

\section{Concluding remarks}

Presently 4-dimensional conformal theories are very scarce and in fact do
not yet exist on the same conceptual and mathematical level as chiral
models. Perturbative ideas about their construction have not yet given
satisfactory results. One idea which was mentioned in the second section as
not leading to physically viable conformal theories was the use of
perturbative Lagrangian AdS\footnote{%
Necessarily with pointlike fields, because there is no known Lagrangian
field theory for extended ``fields''.} models in the $AdS_{5}$-$CQFT_{4}$
connection.

Recently there have also been direct attempts to obtain nontrivial conformal
theories within the family of perturbatively renormalizable supersymmetric
Yang-Mills models with vanishing Callan-Symanzik $\beta $-function (which is
a necessary condition for conformal invariance in perturbation theory) \cite
{hep-th}. There are many lowest order perturbative calculations for gauge
invariant quantities (often involving additional approximations), but they
presently did not reach a level where they could be used for a test with the
braiding ideas in this paper. Even the conjectures on the existence of
certain ``protected'' quantities are not very clear. Protected objects by
definition do not receive any contributions from the interaction i.e. they
retain their zero order free field values. A chiral illustration of a
protected quantity would be a nontrivial chiral theory which has an energy
momentum tensor with the free field value c=1 of which there exist plenty.
``Nontrivial'' here simply means that there are other unprotected quantities
in the model. What is conceivable is a protection of an entire subalgebra as
in this analogy. A protection of only certain correlation functions on the
other hand (say the two-or three-point normalization constant of a certain
operator) which does not follow a (supersymmetric) charge rule which
characterizes a subalgebra of operators is hardly reconcilable with one's
understanding of the omnipresence vacuum polarization in interacting QFT. So
the question of the quality of arguments about protection remains open.

The crucial question for a future comparison with the timelike braid group
structure is whether the perturbation theory can be pushed far enough so
that one can extract anomalous dimensions. If it works it is probably
limited to models with an abelian braid group phase, which in the analogy to
chiral models would mean something analogous to the massless Thirring model
but not to e.g. a minimal model without a coupling deformation parameter.

Since it is not unrealistic to expect that the first conceptually and
mathematically controllable 4-dimensional nontrivial models will be
conformal (because they stay close to free field theories without being
identical to them), this line of research has a certain urgency and
importance for the future of QFT.

Another fascinating but at the same time very speculative idea which emerges
from the present setting on the structure of higher dimensional conformal
theories is the suggestion that there may be charge superselection rules and
inner symmetries which do not respect the inner/spacetime factorization
pattern of the Coleman-Mandula theorem (and a fortiori not its
prerequisites) even after adjusting the prerequisites in order to
incorporate supersymmetry. Braided charges and their fusions certainly have
a very different structure than multiplets of fields on which compact group
act. In fact a timelike braided structure would inexorably remain linked
with spacetime and dynamics. We have gotten so much used to think in terms
of nonabelian symmetry groups that it is helpful to remind oneself from time
to time that there actually does not exist a single exact nonabelian
continuous flavor symmetry in nature. So to look for explanations of the
observed regularities outside group theory may not be that absurd as it
appears at first sight. But the idea that the observed regularities may be
remnants of timelike braidings remains a farfetched wild speculation as long
as there is no understanding of how conformal theories can be naturally
related to particle models in a more controllable way than assigning to them
a short distance universality class..

\textit{Acknowledgements}: I am indebted to Karl-Henning Rehren for making
some suggestions based on an earlier smaller and still somewhat opaque
version. I thank Lucio Fassarel for some mathematical suggestion and for
helping me with the pictures. I owe also thanks to Francesco Toppan for
discussions which were helpful in the formulation. Last not least I would
like to thank Detlev Buchholz for an interesting exchange of emails on the
issue of possible non-locality of timelike-defined double cone algebras.
Through this exchange of mails it became also clear to me that as a result
of his familiarity with zero mass problems the possibility of a timelike
braided structure was probably on his mind some time ago.

\end{document}